\documentclass[10pt,twocolumn]{extarticle}
\usepackage[a4paper,top=19.5mm,bottom=20mm,left=16mm,right=16mm,columnsep=6.3mm,headheight=13pt,headsep=5mm,footskip=10mm]{geometry}
\usepackage[T1]{fontenc}
\usepackage[utf8]{inputenc}
\usepackage{newtxtext,newtxmath}
\usepackage[scaled=.93]{helvet}
\usepackage{microtype}
\usepackage[version=4]{mhchem}
\usepackage{amsmath,graphicx,booktabs,multirow,tabularx,siunitx}
\usepackage[table]{xcolor}
\usepackage{caption,subcaption,titlesec,dblfloatfix,balance,ragged2e,url}
\usepackage{fancyhdr,lastpage}
\usepackage[numbers,sort&compress]{natbib}
\usepackage[hidelinks]{hyperref}
\definecolor{WileyGray}{RGB}{235,235,235}
\titleformat{\section}{\bfseries\fontsize{11.6}{13.4}\selectfont}{\thesection}{.7em}{}
\titlespacing*{\section}{0pt}{12pt}{5pt}
\titleformat{\subsection}{\bfseries\fontsize{10.4}{12.2}\selectfont}{\thesubsection}{.6em}{}
\titlespacing*{\subsection}{0pt}{8pt}{3pt}
\newcommand{\methodhead}[1]{\par\vspace{5pt}\noindent\textit{\textbf{#1:}}\hspace{.35em}}
\DeclareCaptionLabelSeparator{wileysep}{\hspace{4pt}\textbar\hspace{6pt}}
\DeclareCaptionFont{wileylabel}{\bfseries\sffamily\footnotesize}
\fancypagestyle{plain}{\pagestyle{fancy}}
\makeatletter
\newcommand{\WileyGraphic}[2][]{%
 \IfFileExists{#2}{\includegraphics[#1]{#2}}{%
  \fbox{\begin{minipage}[c][42mm][c]{.90\linewidth}\centering\small
  Missing figure asset: \texttt{\detokenize{#2}}\end{minipage}}}}
\makeatother
\begin{document}
\twocolumn[
\begin{@twocolumnfalse}
{\sffamily\footnotesize\bfseries\rule{2.5pt}{10pt}\hspace{6pt}RESEARCH ARTICLE\par}
\vspace{12pt}
{\bfseries\fontsize{18}{21}\selectfont
Attribute-Enriched Crystal Graph Learning for Optical-Spectrum Prediction and Photovoltaic Materials Discovery\par}
\vspace{9pt}
{\fontsize{9.8}{12.2}\selectfont
Quach Thi Thai Binh\textsuperscript{1}, Nguyen Tran Gia Bao\textsuperscript{1},
Vu Thi Hanh Thu\textsuperscript{1,*}, Mingda Li\textsuperscript{2}, and Nguyen Tuan Hung\textsuperscript{3,*}\par}
\vspace{7pt}
{\fontsize{8.5}{10.5}\selectfont
\textsuperscript{1}Faculty of Physics and Physics Engineering, University of Science, Vietnam National University Ho Chi Minh City, Ho Chi Minh City, Viet Nam\\
\textsuperscript{2}Department of Nuclear Science and Engineering, Massachusetts Institute of Technology, Cambridge, Massachusetts, USA\\
\textsuperscript{3}Frontier Research Institute for Interdisciplinary Sciences, Tohoku University, Sendai, Japan\\[3pt]
\textbf{Correspondence:} Vu Thi Hanh Thu (\href{mailto:vththu@hcmus.edu.vn}{vththu@hcmus.edu.vn}); Nguyen Tuan Hung (\href{mailto:nguyen.tuan.hung.e4@tohoku.ac.jp}{nguyen.tuan.hung.e4@tohoku.ac.jp})\par}
\vspace{8pt}
{\fontsize{8.7}{10.7}\selectfont\noindent\textbf{Keywords:} Crystal graph neural networks \textbar{} self-supervised learning \textbar{} dielectric spectra \textbar{} absorption-coefficient prediction \textbar{} photovoltaic screening \textbar{} Zintl-like materials \textbar{} density functional theory\par}
\vspace{7pt}
\noindent\colorbox{WileyGray}{\begin{minipage}{\dimexpr\textwidth-2\fboxsep\relax}
\vspace{5pt}\noindent\textbf{ABSTRACT}\\[3pt]
\fontsize{9.2}{11.4}\selectfont
Accurate prediction of energy-dependent optical responses directly from crystal structures can accelerate the discovery of energy materials. This study develops attribute-enriched Graph Isomorphism Networks with Edge Features (AE-GINE) and a self-supervised variant (SAE-GINE) to predict dielectric spectra from periodic crystal structures. Explicit physicochemical node and edge descriptors distinguish chemically different neighboring atoms with similar geometric separations. Using 9,915 paired crystal-spectrum samples, AE-GINE reduces the test mean absolute error by approximately 10--12\% relative to the basic GINE representation. Self-supervised initialization provides further improvements under selected optical-label conditions. A dedicated SAE-GINE directly predicts absorption-coefficient spectra for photovoltaic screening of 1,172 external Zintl-like and ternary $AB_2C_2$ structures ($C=\mathrm{P}$ or $\mathrm{N}$). The procedure identifies 50 Zintl-like and 16 $AB_2C_2$ candidates above a spectroscopic limited maximum efficiency threshold of 30\%, corresponding to 65 unique structures after deduplication. First-principles optical calculations are used to assess the shortlisted candidates. The framework connects chemically informed crystal representations, optical-spectrum learning, and targeted photovoltaic materials screening.
\vspace{5pt}\end{minipage}}
\vspace{11pt}
\end{@twocolumnfalse}
]
\section{Introduction}
\label{sec:introduction}

Optical properties are central to the design of photovoltaic absorbers, photodetectors, transparent conductors, sensors, and other optoelectronic materials~\cite{Yu2012,Wei2019,kamijo2023touchless}. 
%Energy-dependent optical spectra contain information that cannot be fully represented by scalar quantities such as the electronic band gap, static dielectric constant, absorption onset, or refractive index at a single wavelength \cite{Ibrahim2024,Grunert2024,Woods-Robinson2023}. 
In particular, the real $\varepsilon_1(E)$ and imaginary $\varepsilon_2(E)$ dielectric spectra describe the linear optical response and determine key quantities including the refractive index, extinction coefficient, absorption coefficient, and reflectance~\cite{Wooten1972}. Predicting these responses over a continuous photon-energy range is therefore important for materials whose functionality depends on energy-dependent light-matter interactions~\cite{Hung2024,Grunert2024,Grunert2025}. 

Optical spectra are accessible experimentally through techniques such as spectroscopic ellipsometry and absorption spectroscopy
\cite{gulo2023exploring,gulo2022exploring}. However, high-throughput measurements require suitable samples and remain sensitive to crystal orientation, film thickness, defects, surfaces, and measurement conditions
\cite{Losurdo2009,Potonik2016}. The first-principles calculations based on density functional theory (DFT) provide a consistent computational alternative \cite{Hohenberg1964,Kohn1965}, but DFT requires dense reciprocal-space sampling, sufficient unoccupied electronic states, and explicit evaluation of optical transitions
\cite{gulo2022exploring}. Consequently, dielectric- and
absorption-spectrum calculations are substantially more expensive than many scalar ground-state-property calculations, limiting the availability of consistently labeled crystal-spectrum datasets.

Machine-learning (ML) approaches provide an increasingly attractive route for accelerating optical-property prediction directly from crystal structure \cite{Grunert2024,Grunert2025,Hung2024}. Graph neural networks (GNNs), in particular, represent periodic crystals through atoms and their local environments and demonstrate strong performance across a broad range of
structure-property tasks \cite{Xie2018,Chen2019,Choudhary2021}. Recent studies extend this paradigm from scalar properties to frequency- or energy-dependent optical responses \cite{Hsu2022,Takahashi2026,Hsu2026}. Among them, GNNOpt
establishes a direct structure-to-spectrum framework using an
E(3)-equivariant graph neural network with optimized atomic embeddings and
demonstrates that learned optical responses can support photovoltaic screening
followed by first-principles validation \cite{Hung2024}. 
%GNNOpt therefore provides an important benchmark for structure-based optical-spectrum learning and establishes the feasibility of connecting crystal-graph prediction with application-oriented materials screening.
An important remaining question concerning these ML models is the
information supplied to the crystal-graph representation itself. This issue is
general to many GNN-based materials approaches rather than specific to a
particular architecture. In many crystal-graph formulations, atomic identity
and physicochemical descriptors are primarily associated with nodes, whereas
relationships between neighboring atoms are represented mainly through
geometric quantities such as distances, relative positions, displacement
vectors, or radial and angular basis functions
\cite{Xie2018,Chen2019,Choudhary2021,Hung2024}. These formulations provide
effective descriptions of local geometry, but explicit physicochemical
contrasts between connected atoms are less commonly exposed as dedicated edge
attributes. As a result, atomic pairs with similar geometric separations can
remain similar from a geometry-centered representation despite substantial
differences in electronegativity, atomic size, ionization tendency, or
periodic-table identity.

This observation motivates an attribute-enriched representation in which
chemical information is made explicitly available at both the node and edge
levels. In addition to atomic descriptors and interatomic geometry, pairwise
physicochemical contrasts between connected atoms participate directly in
edge-aware message passing. The resulting scientific question is therefore
whether explicit node-edge enrichment provides a measurable benefit for
energy-resolved optical learning beyond a simpler graph representation. We
evaluate this contribution by comparing the complete attribute-enriched
representation with the basic GINE (Graph Isomorphism Network with Edge Features~\cite{hu2019strategies}) representation within the same structure-to-spectrum framework, without assigning the resulting improvement
to any single descriptor.

A second challenge concerns optical-label efficiency. Optical-spectrum learning
requires crystal structures paired with computationally expensive spectral
labels, although crystal structures themselves can be available without
corresponding optical calculations
\cite{Grunert2024,Hung2024,Jain2013}. Self-supervised learning provides a
complementary route in which the graph encoder first learns structure-informed
representations before downstream optical supervision
\cite{Magar2022,Fu2024}. This strategy is particularly relevant to
energy-resolved spectra because generating first-principles spectral labels is
substantially more expensive than obtaining the underlying crystal structures.
Although self-supervised crystal-graph learning demonstrates useful
representations for materials-property prediction
\cite{Magar2022,Fu2024}, its contribution to full dielectric-spectrum learning,
particularly under systematically reduced optical-label availability, remains
insufficiently established.

%Building on our previous GNNOpt framework \cite{Hung2024}, the present study therefore focuses on two complementary advances in crystal representation and label efficiency. First, we investigate whether explicit physicochemical information at atomic nodes and interatomic edges improves energy-resolved optical learning. Second, we examine whether self-supervised pretraining provides an additional benefit when optical-spectrum labels are limited. The advance therefore does not arise simply from increasing the labeled data volume: attribute enrichment determines which chemical information is explicitly available to the graph network, whereas self-supervised pretraining determines how this information is organized before optical-spectrum supervision. The present work consequently addresses broader representation-and label-efficiency questions in crystal GNNs rather than a limitation specific to any previous architecture.

To address these challenges, we develop two attribute-enriched variants of the
GINE for direct prediction of $\varepsilon_1(E)$ and $\varepsilon_2(E)$ from periodic crystal structures. The randomly initialized attribute-enriched model is called \textbf{AE-GINE} (Attribute-Enriched GINE), whereas its self-supervised counterpart is called \textbf{SAE-GINE} (Self-Supervised Attribute-Enriched GINE). The term GINE refers to the basic GINE baseline unless otherwise specified. The attribute-enriched representation combines chemically informative atomic descriptors with geometric features and explicit pairwise physicochemical edge attributes, allowing node-level chemistry and neighboring-atom contrasts to participate directly in edge-aware message passing. For SAE-GINE, the same encoder is pretrained without dielectric-spectrum labels through topology-preserving feature corruption and reconstruction of structure-derived graph statistics before supervised spectrum learning. AE-GINE is compared with the basic GINE representation to evaluate attribute enrichment, whereas SAE-GINE is compared with AE-GINE under strictly matched full-data and reduced-label conditions to isolate the contribution of self-supervised initialization.

%Beyond dielectric-spectrum prediction, we examine whether the learned crystal representation can support photovoltaic candidate prioritization. The spectroscopic limited maximum efficiency (SLME) depends explicitly on the energy-dependent absorption coefficient, $\alpha(E)$ \cite{Yu2012}. Previous optical-GNN results suggest that direct prediction of a response-specific optical target can be advantageous relative to obtaining the same response through transformations of separately predicted dielectric spectra \cite{Hung2024}. We therefore formulate absorption as a dedicated prediction task and train an SAE-GINE model to predict $\alpha(E)$ directly from crystal structure. The predicted absorption spectra are subsequently used to calculate SLME, thereby connecting the learned crystal representation directly to the optical response used for photovoltaic prioritization.

%The SLME analysis is performed under a common radiative-limit screening protocol with the radiative fraction fixed at $f_r=1$ \cite{Yu2012}. Material-specific non-radiative recombination kinetics are not included in the present ranking. Practical photovoltaic performance also depends on carrier recombination, transport, defects, interfaces, and device architecture; accordingly, SLME is used here as a consistent optical-screening metric for candidate prioritization rather than as a complete prediction of device efficiency.

Beyond dielectric-spectrum prediction, we examine whether the learned crystal representation can support photovoltaic candidate prioritization.
To test this screening strategy in a physically motivated materials space, we
focus on Zintl-like compounds together with ternary $AB_2C_2$ compounds with
$C=\mathrm{P}$ or $\mathrm{N}$. The choice of the Zintl-related space is
supported by experimental and computational evidence that representative Zintl
phosphides can exhibit photovoltaic-relevant band gaps, strong optical
absorption, favorable carrier properties, and low cost compared with silicon-based or perovskite materials~\cite{Yuan2024}. For example, CaZn$_2$P$_2$ is
experimentally demonstrated as a stable Zintl-phosphide thin-film
semiconductor with strong visible-light absorption, a direct optical gap near
1.95~eV, and measurable photoexcited-carrier lifetimes, making it a candidate
top absorber for tandem photovoltaic applications \cite{Quadir2024}.
BaCd$_2$P$_2$ is independently identified through high-throughput
first-principles screening and experimental characterization as a stable solar
absorber with strong optical absorption and long carrier lifetime
\cite{Yuan2024}. First-principles studies of K$_3$Ag$_3$As$_2$ further report
a direct semiconducting gap and optical response relevant to photovoltaic and
optoelectronic applications \cite{Kimuyu2024}. These examples establish the
broader Zintl-related chemical space as a physically motivated test bed rather
than an arbitrary application of a general optical predictor. Using SAE-GINE for photovoltaic screening of 1,172 external Zintl-like and ternary \(AB_2C_2\) structures with \(C=\mathrm{P}\) or \(\mathrm{N}\), we identify 50 Zintl-like and 16 \(AB_2C_2\) shortlisted candidates with predicted spectroscopic limited maximum efficiency (SLME) above 30\%, corresponding to 65 unique structures after cross-pool deduplication. A subset of 40 candidates from the Zintl-like
shortlist is subsequently evaluated using DFT optical calculations.

\section{Attribute-Enriched Crystal Graph Learning for Dielectric-Spectrum Prediction}
\label{sec:graph_prediction}

\begin{figure*}[!t]
    \centering
    \WileyGraphic[width=0.95\textwidth]{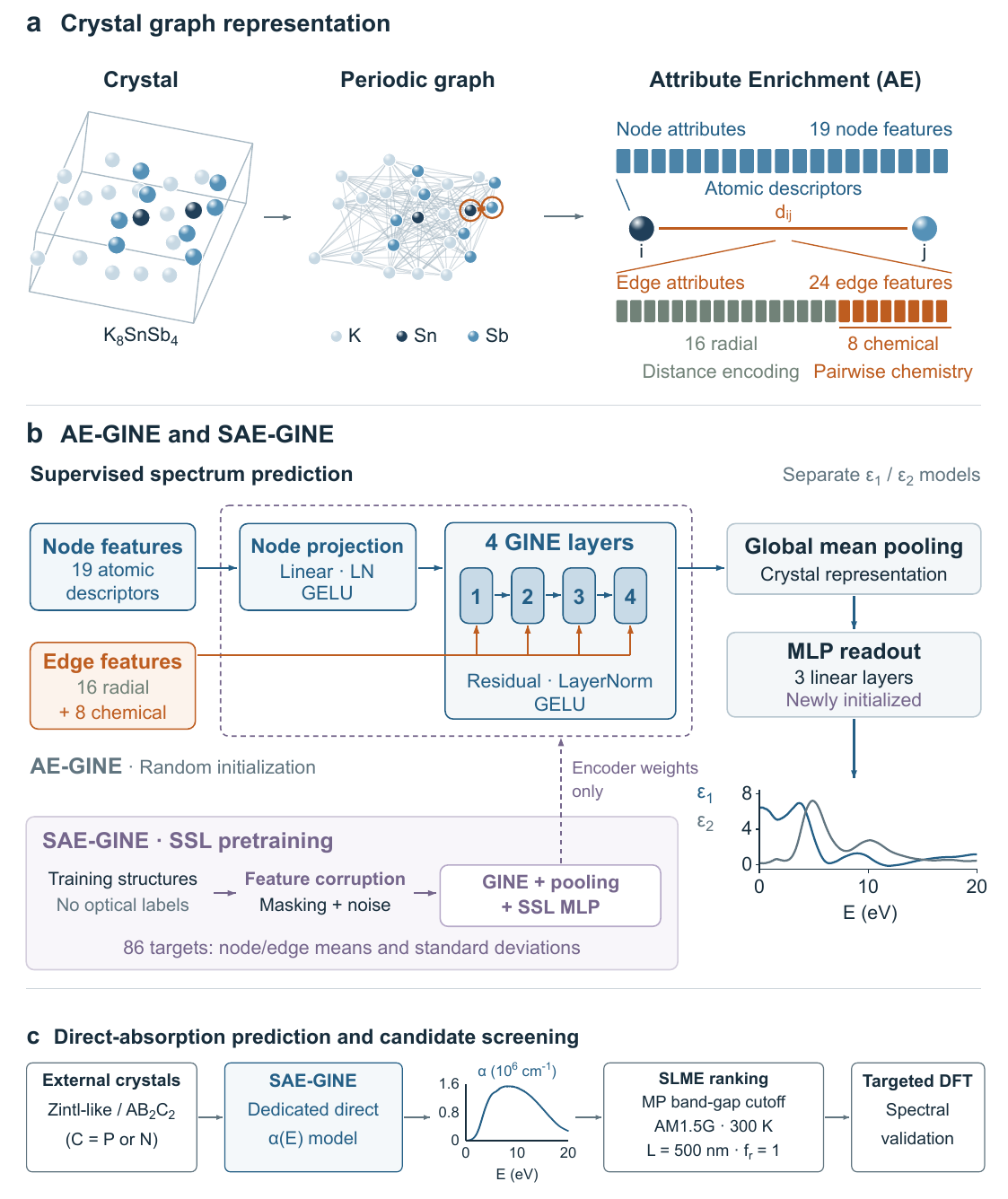}
    \caption{
    Overview of the self-supervised, attribute-enriched crystal-graph framework for
    dielectric-spectrum prediction and photovoltaic screening.
    (a) Periodic crystal structures are converted into graphs containing chemically
    informative node attributes, radial distance features, and explicit pairwise
    physicochemical edge attributes. The primary graph uses 19 node-features
    and 24 edge-features, whereas the 20-node-features/24-edge-features variant
    is retained only as an auxiliary configuration.
    (b) The GINE backbone is pretrained through node masking and Gaussian feature
    corruption while preserving graph topology, with graph-level node and edge
    descriptor statistics used as reconstruction targets. The pretrained backbone is
    subsequently transferred to separate predictors for \(\varepsilon_1(E)\) and
    \(\varepsilon_2(E)\), whose spectral prediction heads are newly initialized.
    Label efficiency is evaluated using \(5\%\), \(10\%\), \(25\%\), and \(50\%\)
    of the labeled training data, together with a full-data benchmark.
    (c) For photovoltaic screening, a SAE-GINE model directly predicts the absorption-coefficient spectrum, \(\alpha(E)\), from crystal structure. The predicted spectra are used to calculate the spectroscopic limited maximum efficiency (SLME) at a thickness of 500~nm, rank external Zintl-like and \(AB_2C_2\) candidates with \(C=\mathrm{P}\) or \(\mathrm{N}\), remove structures overlapping with the original database, and prioritize candidates for first-principles validation.
    }
    \label{fig:framework}
\end{figure*}

\begin{table*}[!t]
    \centering
    \caption{
    Test performance for predicting the real and imaginary dielectric spectra.
    MAE values are reported as the mean \(\pm\) sample standard deviation over
    three independent runs. The rows represent the final evaluated configurations
    and should not be interpreted as a strictly hyperparameter-controlled
    comparison of initialization strategies.
    }
    \label{tab:main_benchmark}

    {\small
    \renewcommand{\arraystretch}{1.12}
    \setlength{\tabcolsep}{5.5pt}

    \begin{tabular*}{\textwidth}{@{\extracolsep{\fill}}lllcc@{}}
        \toprule
        Model &
        Graph representation &
        Initialization &
        MAE, \(\varepsilon_1(E)\) &
        MAE, \(\varepsilon_2(E)\) \\
        \midrule

        GCN &
        Basic &
        Random &
        \(0.460\pm0.007\) &
        \(0.492\pm0.010\) \\

        GraphSAGE &
        Basic &
        Random &
        \(0.395\pm0.006\) &
        \(0.403\pm0.006\) \\

        GINE &
        Basic &
        Random &
        \(0.316\pm0.013\) &
        \(0.322\pm0.008\) \\

        AE-GINE &
        Attribute-enriched &
        Random &
        \(\mathbf{0.277\pm0.004}\) &
        \(0.289\pm0.014\) \\

        SAE-GINE &
        Attribute-enriched &
        Self-supervised &
        \(0.282\pm0.005\) &
        \(\mathbf{0.286\pm0.005}\) \\

        \bottomrule
    \end{tabular*}
    }
\end{table*}

\begin{figure*}[!t]
    \centering
    \WileyGraphic[width=0.96\textwidth]{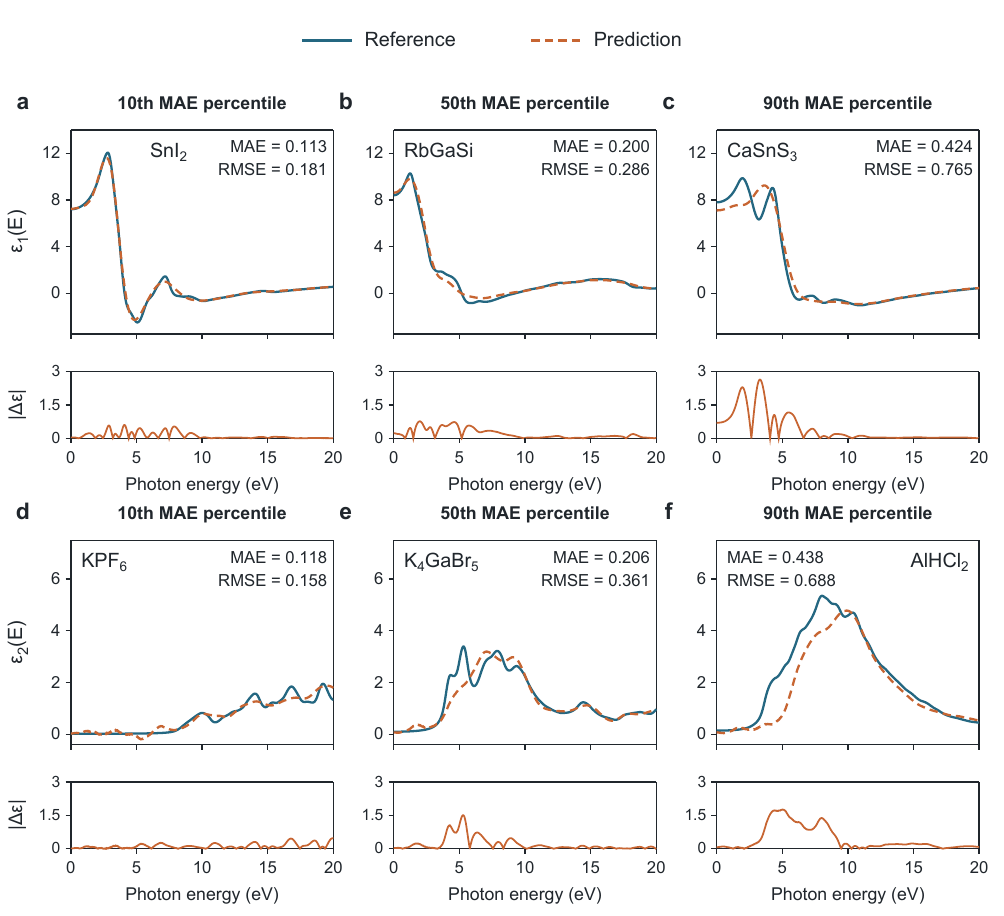}
    \caption{
    Representative held-out dielectric spectra and prediction errors.
    (a-c) Real dielectric spectra from the three-seed AE-GINE ensemble;
    (d-f) imaginary dielectric spectra from the three-seed SAE-GINE
    ensemble. For each target, examples were selected nearest to the
    10th, 50th, and 90th percentiles of the per-structure ensemble mean
    absolute error (MAE) distribution. Solid blue and dashed orange
    curves denote reference and predicted spectra, respectively.
    Vertical spectral limits are identical within each row.
    Lower traces show pointwise absolute errors,
    $|\Delta\varepsilon(E)| =
    |\varepsilon_{\mathrm{pred}}(E)-\varepsilon_{\mathrm{ref}}(E)|$,
    on a common vertical scale across all six examples.
    MAE and root mean squared error (RMSE) are computed over the same
    2001 energy points from 0 to 20~eV. RMSE is a complementary
    diagnostic and was not used to select examples.
    The two rows contain independently selected structures,
    not paired materials.
    }
    \label{fig:representative_spectra}
\end{figure*}

In Figure~\ref{fig:framework}, we show the framework from periodic crystal structures to dielectric-spectrum prediction and ML-guided photovoltaic screening. As shown in Figure~\ref{fig:framework}a, each crystal structure is represented as a periodic graph in which atomic sites form the nodes and neighboring atomic pairs form the edges. The primary attribute-enriched representation contains a 19-dimensional node feature vector and a 24-dimensional edge feature vector (see detailed features in Tables S2 and S3 of the Supplementary Information). The node features describe elemental and physicochemical information, including periodic-table position, electronegativity, atomic size, ionization tendency, valence-shell occupation, and metallic character. The edge features combine 16 radial-basis distance channels with eight explicit pairwise attributes that encode chemical and periodic-table contrasts between connected atoms. Consequently, neighboring atomic pairs with similar geometric separations need not share the same edge representation when their physicochemical identities differ. The enriched graph therefore exposes both local geometry and pairwise chemical contrast directly to the GINE message-passing operation, so that chemical information is represented explicitly at both the atomic-node and interatomic-edge levels. 

For the AE-GINE model, the enriched graph is processed by a four-layer GINE encoder that incorporates edge attributes directly into message passing, as shown in Figure~\ref{fig:framework}b. Global mean pooling produces a crystal-level
representation, which is mapped to a 2,001-point dielectric spectrum on the common
0-20~eV energy grid. Separate models are trained for the real and imaginary dielectric responses, \(\varepsilon_1(E)\) and \(\varepsilon_2(E)\), respectively. This design allows the model to learn energy-dependent optical responses from a crystal-level representation while retaining both atomic and pairwise physicochemical information throughout message passing.

For the SAE-GINE model, the same GINE backbone is pretrained without dielectric-spectrum labels, as shown in Figure~\ref{fig:framework}b. Node masking and Gaussian feature corruption are applied to the graph attributes while the graph connectivity remains unchanged. The encoder is optimized to reconstruct graph-level statistics derived from the original node and edge descriptors. After pretraining, the GINE backbone is transferred to the downstream dielectric models, whereas the spectral prediction heads are newly initialized. The self-supervised stage therefore organizes the chemical, geometric, and pairwise information already encoded in the enriched graph into a structure-informed graph representation before optical supervision is introduced. The downstream stage then learns the mapping from this pretrained representation to the dielectric spectra.

Table~\ref{tab:main_benchmark} compares the final evaluated graph-learning
configurations, including three basic graph representations (GCN, GraphSAGE, and GINE) and two attribute-enriched representations (AE-GINE and SAE-GINE) on the fixed held-out test set. For the basic graph representation, GINE produces substantially lower errors than the evaluated GCN and GraphSAGE configurations for both $\varepsilon_1(E)$ and $\varepsilon_2(E)$. This is because the implemented GCN and GraphSAGE models do not incorporate the stored edge attributes in the same way as GINE does. For attribute-enriched representations, AE-GINE and SAE-GINE show improvements on the final full-data benchmark.
%, these results compare the complete evaluated configurations rather than constituting a strictly controlled message-passing-operator ablation.
%Within the GINE architecture, the attribute-enriched representation provides the largest improvement in the final full-data benchmark. 
Compared with the original GINE baseline, AE-GINE reduces the
$\varepsilon_1(E)$ MAE from 0.316 to 0.277 and the
$\varepsilon_2(E)$ MAE from 0.322 to 0.289, corresponding to relative reductions of 12.3\% and 10.2\%, respectively. SAE-GINE performance is also close to that of AE-GINE, with MAEs of 0.282 and 0.286 for $\varepsilon_1(E)$ and $\varepsilon_2(E)$, respectively. These results therefore suggest that enriching the crystal graph beyond a basic atomic-geometric representation improves full dielectric-spectrum learning under the evaluated GINE configuration. 

To further examine the relative weighting of the enriched descriptors, we employ an auxiliary 20-node/24-edge linear-ensemble model, in which the node representation includes an additional covalent-radius descriptor used only for this auxiliary analysis. Each scalar node or edge descriptor is projected through an independent learnable branch, and the resulting representations are combined using softmax-normalized mixing coefficients
that sum to unity separately within the node and edge feature families. Separate coefficient sets are learned for $\varepsilon_1(E)$ and $\varepsilon_2(E)$ and evaluated over three independent training seeds. The complete node- and edge-feature mixing-probability profiles are presented in Figures S1 and S2, respectively, with the corresponding seed-resolved values reported in Tables S4, S5, S6, and S7 of the Supporting Information. It is noted that these coefficients are specific to the auxiliary analysis and are not part of the architecture, training, or inference of the primary 19-node/24-edge AE-GINE or SAE-GINE models.

%This comparison evaluates the complete graph representations rather than the effect of any single descriptor. The basic and attribute-enriched graphs differ jointly in descriptor content, dimensionality, scaling, missing-value treatment, and pairwise chemical information. The results therefore show that the enriched representation improves the overall structure-to-spectrum mapping under the evaluated configuration, without assigning the improvement to an individual atomic or pairwise chemical feature.

%The SAE-GINE in Table~\ref{tab:main_benchmark} represents the
%final evaluated self-supervised configuration and is not the strictly matched initialization experiment. Its full-data performance remains close to that of AE-GINE, with MAEs of 0.282 and 0.286 for \(\varepsilon_1(E)\) and \(\varepsilon_2(E)\), respectively. The effect of self-supervised initialization is therefore evaluated separately under matched full-data and reduced-label conditions, for which encoder initialization is the controlled variable.

In Figure~\ref{fig:representative_spectra}, we show the representative held-out dielectric spectra and prediction errors. For each dielectric component, the configuration used for the representative spectra is selected using the three-run mean validation MAE without reference to held-out test performance. This procedure selects AE-GINE for \(\varepsilon_1(E)\) and SAE-GINE for \(\varepsilon_2(E)\). Structures are selected near the 10th, 50th, and 90th percentiles of the corresponding per-structure ensemble-MAE distributions. This percentile-based selection provides examples that span the held-out test-error distribution, rather than selecting spectra based on visual agreement. The selected \(\varepsilon_1(E)\) examples have MAEs of 0.113, 0.200, and 0.424, whereas the corresponding \(\varepsilon_2(E)\) examples have MAEs of 0.118, 0.206,
and 0.438. The percentile-selected examples show that the models retain the broad
spectral envelopes and dominant response regions more consistently than narrow peaks,
closely spaced features, and precise response onsets. In the illustrated higher-error
cases, the dominant deviations arise from localized spectral features, peak positions,
and response onsets rather than from a complete loss of the broad dielectric response.

Together, these results establish the dielectric-spectrum prediction performance and
show that explicit enrichment of the crystal-graph representation improves the
structure-to-spectrum mapping within the evaluated GINE framework. The role of
self-supervised initialization is examined next under strictly matched full-data and
reduced-label conditions, separating representation enrichment from the contribution
of pretraining without optical-spectrum labels. The subsequent photovoltaic analysis then treats absorption as a distinct response-specific prediction task, in which a dedicated SAE-GINE directly predicts \(\alpha(E)\) for SLME-guided candidate prioritization and first-principles validation (see Theoretical Section).

\section{Self-Supervised Initialization and Label Efficiency}
\label{sec:ssl_label_efficiency}

To isolate the effect of encoder initialization, we perform a strictly matched comparison in which AE-GINE and SAE-GINE use identical crystal graphs, fixed training, validation, and test partitions, network architectures, optimization parameters, and training seeds. The only controlled difference is whether
the GINE backbone is initialized randomly or from the self-supervised checkpoint.
Self-supervised pretraining uses only structures from the training partition and does not access the validation or held-out test structures, thereby preventing information
leakage across the fixed data split.

As summarized in Table~\ref{tab:matched_full_data}, self-supervised initialization reduces the three-run mean test MAE from \(0.2881\pm0.0079\) to \(0.2808\pm0.0021\) for \(\varepsilon_1(E)\), corresponding to a relative reduction
of 2.5\%. For \(\varepsilon_2(E)\), the mean MAE decreases from
\(0.2973\pm0.0077\) to \(0.2816\pm0.0039\), corresponding to a relative reduction of 5.3\%.
%The improvement occurs in two of the three \(\varepsilon_1(E)\) runs and in all three \(\varepsilon_2(E)\) runs. 
Self-supervised initialization therefore improves the mean
full-data performance for both dielectric responses, with a larger and more consistent
effect for \(\varepsilon_2(E)\). Importantly, this matched experiment isolates encoder
initialization as the controlled variable. It is therefore distinct from
Table~\ref{tab:main_benchmark}, which summarizes the final evaluated configurations
rather than a strictly matched initialization comparison.

\begin{table}[!t]
    \centering
    \caption{
    Strictly matched full-data comparison between AE-GINE and SAE-GINE. Both models use the same attribute-enriched graph representation, fixed data partition, network architecture, optimization settings, and training seeds; encoder initialization is the only controlled difference. Values are reported as the mean \(\pm\) sample standard deviation over seeds 42, 123, and 2025.
    }
    \label{tab:matched_full_data}
    {\small
    \begin{tabular}{lccc}
        \toprule
        \(\varepsilon(E)\) & AE-GINE & SAE-GINE & MAE reduction \\
        \midrule
        \(\varepsilon_1(E)\) &
        \(0.2881\pm0.0079\) &
        \(\mathbf{0.2808\pm0.0021}\) &
        2.5\% \\
        \(\varepsilon_2(E)\) &
        \(0.2973\pm0.0077\) &
        \(\mathbf{0.2816\pm0.0039}\) &
        5.3\%  \\
        \bottomrule
    \end{tabular}
    }
\end{table}

\begin{figure*}[!t]
    \centering
    \WileyGraphic[
        width=0.84\textwidth
    ]{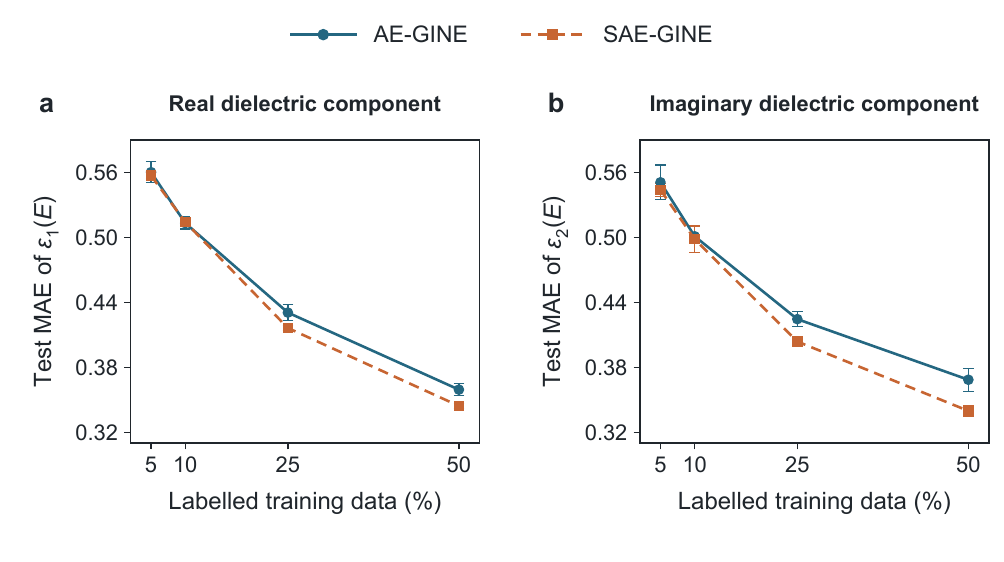}
    \caption{
    Label-efficiency analysis of self-supervised initialization for dielectric-spectrum
    prediction.
    (a) Test MAE for \(\varepsilon_1(E)\) and
    (b) test MAE for \(\varepsilon_2(E)\) using \(5\%\), \(10\%\), \(25\%\), and
    \(50\%\) of the labeled training data.
    AE-GINE and SAE-GINE are evaluated using identical labeled subsets,
    graph representations, optimization settings, validation partitions, and test sets.
    Each point represents the mean over three training seeds, and the error bars denote
    one sample standard deviation.
    }
    \label{fig:limited_label}
\end{figure*}

In Figure~\ref{fig:limited_label}, we show the label efficiency of the SAE-GINE compared with that of the AE-GINE model. Only \(5\%\), \(10\%\), \(25\%\), or \(50\%\) of the
labeled training structures are retained, while AE-GINE and SAE-GINE use
identical labeled subsets, attribute-enriched graph representations, optimization
settings, validation partitions, and test sets. The self-supervised checkpoint is
pretrained using crystal graphs from the full fixed training partition without using
their dielectric-spectrum labels, whereas the label fractions refer only to the
structures whose dielectric spectra are used during downstream supervised training.
The reduced-label experiments therefore evaluate optical-label efficiency under fixed
access to the training-partition crystal structures rather than reduced access to
structural information.

Here, we define the relative MAE change, $\Delta_{\mathrm{MAE}}$, as follows:
\begin{equation}
\Delta_{\mathrm{MAE}}
=
\frac{
\mathrm{MAE}_{\mathrm{AE-GINE}}
-
\mathrm{MAE}_{\mathrm{SAE-GINE}}
}{
\mathrm{MAE}_{\mathrm{AE-GINE}}
}
\times 100\%,
\end{equation}
in which the positive values indicate an improvement from self-supervised initialization and the negative values indicate a degradation. For \(\varepsilon_1(E)\), $\Delta_{\mathrm{MAE}}$ associated with self-supervised
initialization are 0.6\%, \(-0.2\%\), 3.3\%, and 4.2\% at the \(5\%\), \(10\%\),
\(25\%\), and \(50\%\) labeled fractions, respectively. The differences at \(5\%\)
and \(10\%\) remain negligible, whereas clearer improvements emerge at \(25\%\) and
\(50\%\). At these two fractions, the mean MAE decreases from 0.4307 to 0.4165 and
from 0.3595 to 0.3444, respectively.
The limited-label benefit is stronger for \(\varepsilon_2(E)\). Self-supervised
initialization reduces the mean MAE by 1.3\%, 0.5\%, 4.9\%, and 7.8\% across the
\(5\%\), \(10\%\), \(25\%\), and \(50\%\) labeled fractions, respectively.
The largest reduction occurs at the \(50\%\) fraction, where the mean MAE decreases
from 0.3687 to 0.3399.

Overall, self-supervised initialization does not yield uniform gains across all evaluated
label fractions. Its effect remains small at the \(5\%\) and \(10\%\) fractions,
whereas clearer improvements emerge at \(25\%\) and \(50\%\), particularly for
\(\varepsilon_2(E)\). Together with the representation comparison in
Table~\ref{tab:main_benchmark}, these results indicate complementary roles for explicit
graph enrichment and self-supervised pretraining. The attribute-enriched graph defines
the chemical, geometric, and pairwise information available to the encoder, whereas
self-supervised pretraining organizes this information before optical-spectrum
supervision is introduced. The resulting benefit is most evident in the matched
full-data comparison and at intermediate levels of optical-label availability rather
than uniformly across all evaluated label fractions.

\section{Kramers--Kr\"onig Relations and SLME Validation}
\label{sec:optical_ranking}

Since the two dielectric components are predicted independently, their mutual physical consistency is additionally examined using a post-hoc finite-window Kramers-Kr\"onig (KK) diagnostic. By using Equations~\eqref{eq:kk1} and~\eqref{eq:kk2}, the prediction-level KK residuals exceed the corresponding finite-window reference baselines by 49.9\% for 
\(\varepsilon_2\!\rightarrow\!\varepsilon_1^{\mathrm{KK}}\) and 42.4\% for
\(\varepsilon_1\!\rightarrow\!\varepsilon_2^{\mathrm{KK}}\), while the median
spectral Pearson coefficients remain 0.997 and 0.993, respectively. The independently predicted components therefore retain strong spectral-shape correspondence but do not satisfy strict quantitative Kramers-Kr\"onig consistency. Detailed residual and window-sensitivity analyses are provided in Figure S3 of the Supporting Information.

\begin{figure}[!t]
    \centering
    \WileyGraphic[width=0.95\columnwidth]
    {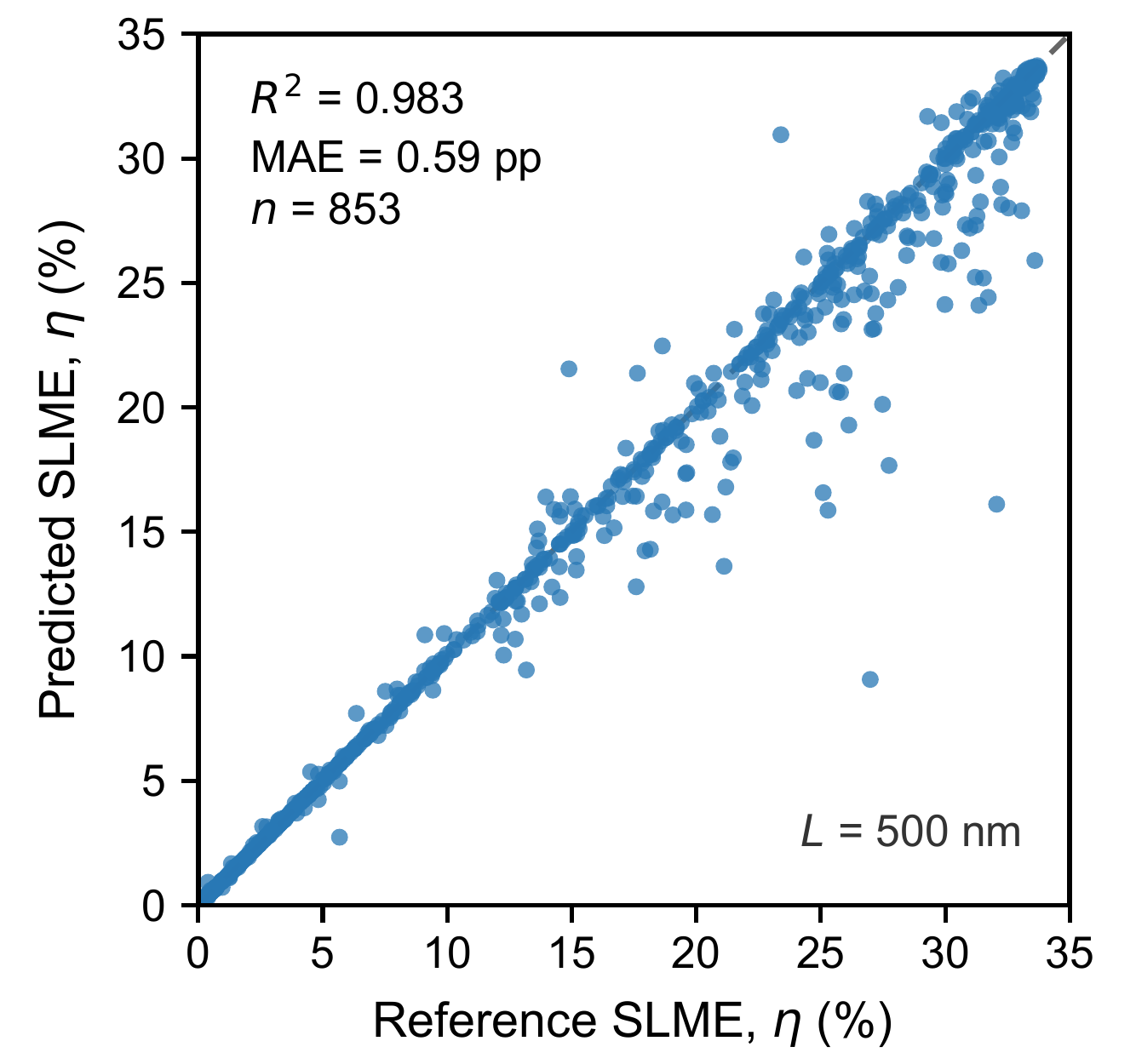}
    \caption{
    %Held-out validation of SLME calculated from directly predicted absorption spectra.
    Comparison between predicted and reference SLME values for the 853 cases with finite SLME values in the 992-structure held-out test set by using SAE-GINE.
    %(b) Overlap between predicted and reference top-\(k\) SLME-ranked materials for
    %\(k=10\), 20, 50, and 100. The predicted ranking retains 5 of the reference top 10,
    %13 of the top 20, 44 of the top 50, and 89 of the top 100 materials.
    }
    \label{fig:slme_validation}
\end{figure}

For photovoltaic evaluation, the absorption-coefficient spectrum
$\alpha(E)$ is treated as a dedicated response-specific prediction task.
Although $\alpha(E)$ can be derived from predicted $\varepsilon_1(E)$ and
$\varepsilon_2(E)$~\cite{Hung2024}, this route can propagate and combine
errors from the independently predicted dielectric components into the
derived absorption response. Because the spectroscopic limited maximum
efficiency (SLME) depends explicitly on the energy-dependent absorption
coefficient, an optimized SAE-GINE predicts $\alpha(E)$ directly from
crystal structure. This response-specific formulation is also motivated
by previous optical-GNN studies showing that direct prediction of an
optical response can be advantageous relative to deriving it from
separately predicted dielectric components~\cite{Hung2024}.

To evaluate this choice directly in the present framework, a matched
held-out comparison between the dielectric-derived route
($\varepsilon_1,\varepsilon_2 \rightarrow \alpha \rightarrow \mathrm{SLME}$)
and the direct-absorption route
($\alpha \rightarrow \mathrm{SLME}$) is provided in Figure~S4 of the Supporting Information. The results show that $\alpha \rightarrow \mathrm{SLME}$ ($R^2=0.983$) is slightly better than $\varepsilon_1,\varepsilon_2 \rightarrow \alpha \rightarrow \mathrm{SLME}$ ($R^2=0.982$) on the same 853 finite test cases. Therefore, we use the direct-$\alpha(E)$ SAE-GINE model for the subsequent screening workflow.

The predicted absorption spectra are used to calculate SLME through the detailed-balance formulation in Equations~\eqref{eq:slme_absorptivity}-\eqref{eq:slme}. For the held-out comparison, the absorption coefficient below the independently available direct band gap is set to zero for both the reference and predicted spectra before SLME evaluation. Applying the same gap cutoff to both spectra ensures that the comparison isolates differences in the absorption response rather than differences arising from band-gap treatment.

\begin{figure*}[!th]
    \centering
    \WileyGraphic[width=0.96\textwidth]{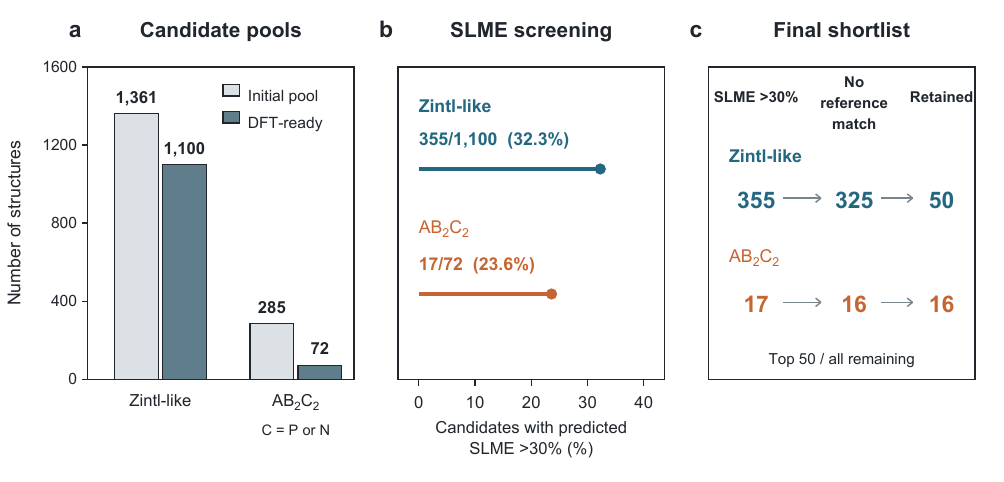}
    \caption{
    External screening and construction of the final candidate shortlist.
    (a) Initial and DFT-ready candidate pools. DFT-ready structures satisfy an MP band gap of 0.30-4.50~eV and an energy above the hull of $\leq 0.10$~eV/atom before optical screening.
    (b) Fractions of DFT-ready Zintl-like and $AB_2C_2$ structures with predicted SLME $>30\%$.
    (c) Candidate reduction after reference-database structure-overlap filtering and final shortlist selection. Of the 355 Zintl-like and 17 $AB_2C_2$ candidates above the SLME threshold, 325 and 16, respectively, remain without a reference-structure match. The 50 highest-ranked Zintl-like candidates and all 16 remaining $AB_2C_2$ candidates are retained. One structure, Ba(GaP)$_2$ (mp-ckxsa), occurs in both retained pools, yielding 65 unique candidates after cross-pool deduplication. Structural novelty is defined relative to the adopted reference database and matching protocol.
    }
    \label{fig:external_screening}
\end{figure*}

In Figure~\ref{fig:slme_validation}, we compare the resulting SLME values and candidate rankings obtained from the predicted and reference absorption spectra. The predicted SLME values agree closely with the reference values, with Pearson and Spearman correlation coefficients of 0.992 and 0.993, respectively, \(R^2=0.983\), and an MAE of 0.59 percentage points. The maximum reference and predicted SLME values are 33.77\% and 33.72\%, respectively. Compared with the GNNOpt~\cite{Hung2024} model (with $R^2=0.81$ on a dataset of 944 materials), which uses only attribute-enriched node representations, the SAE-GINE shows an improvement due to both attribute-enriched node and edge representations, as well as a larger dataset (9,915 materials) for training.

\section{External SLME Screening and Candidate Prioritization}
\label{sec:external_screening}

The direct-\(\alpha(E)\) SAE-GINE is applied to two external candidate spaces: a compositionally defined Zintl-like pool and a ternary \(AB_2C_2\) pool with
\(C=\mathrm{P}\) or \(\mathrm{N}\). As mentioned in the introduction, we focus on Zintl-like and ternary \(AB_2C_2\) materials due to recent interest in their applications in solar cells~\cite{Quadir2024,Yuan2024}. It notes that the direct-\(\alpha(E)\) SAE-GINE model can be extended to apply to various photovoltaic material families, such as perovskites.

The initial searches for Zintl-like and ternary \(AB_2C_2\) materials contain 1,361 and 285 structures, respectively. After applying the electronic-structure and stability criteria described in the Theoretical Section, 1,100 Zintl-like and 72 \(AB_2C_2\) structures remain, giving 1,172 external candidates for SLME screening. Here, Zintl-like denotes the composition-based screening rule and does not imply a confirmed Zintl-phase assignment.

For each eligible structure, the three-seed SAE-GINE ensemble directly predicts
the absorption-coefficient spectrum, \(\alpha(E)\). SLME is then evaluated using \(L=500~\mathrm{nm}\), \(T=300~\mathrm{K}\), and \(f_{\mathrm r}=1\).
Because independent direct-gap labels are unavailable for the external candidate pools, the Materials Project band gap is used as a common numerical cutoff when evaluating the predicted absorption spectra.

As summarized in Figure~\ref{fig:external_screening}, 355 of the 1,100
Zintl-like structures and 17 of the 72 \(AB_2C_2\) structures satisfy the
primary criterion of \(\mathrm{SLME}>30\%\), corresponding to retention
fractions of 32.3\% and 23.6\%, respectively. Structure-level overlap with the
original crystal-spectrum database is then removed according to the procedure
described in the Experimental Section. This step leaves 325 Zintl-like
candidates and 16 \(AB_2C_2\) candidates that do not overlap with the original
crystal-spectrum database. The Zintl-like set is subsequently reduced to the
50 highest-ranked candidates according to predicted SLME, whereas all 16
remaining \(AB_2C_2\) candidates are retained. The two source-pool shortlists
therefore contain 66 memberships. Because one structure occurs in both source
pools, the combined shortlist contains 65 unique candidates for subsequent
first-principles evaluation. The complete 65-material portfolio and its
source-pool provenance are reported in Table S8 of the Supporting Information,
while the final 50-material Zintl-like shortlist, including its DFT-validation
status, is provided in Table S9.

The predicted SLME values of the 65 unique candidates span 30.16-33.75\%.
These structures constitute a pre-DFT shortlist rather than confirmed
photovoltaic materials. The external-screening workflow therefore serves as a
ranking-and-reduction step that reduces the 1,172-structure search space to a
tractable set of high-priority candidates for targeted first-principles
evaluation.

\begin{figure*}[!t]
    \centering
    \WileyGraphic[width=0.95\textwidth]
    {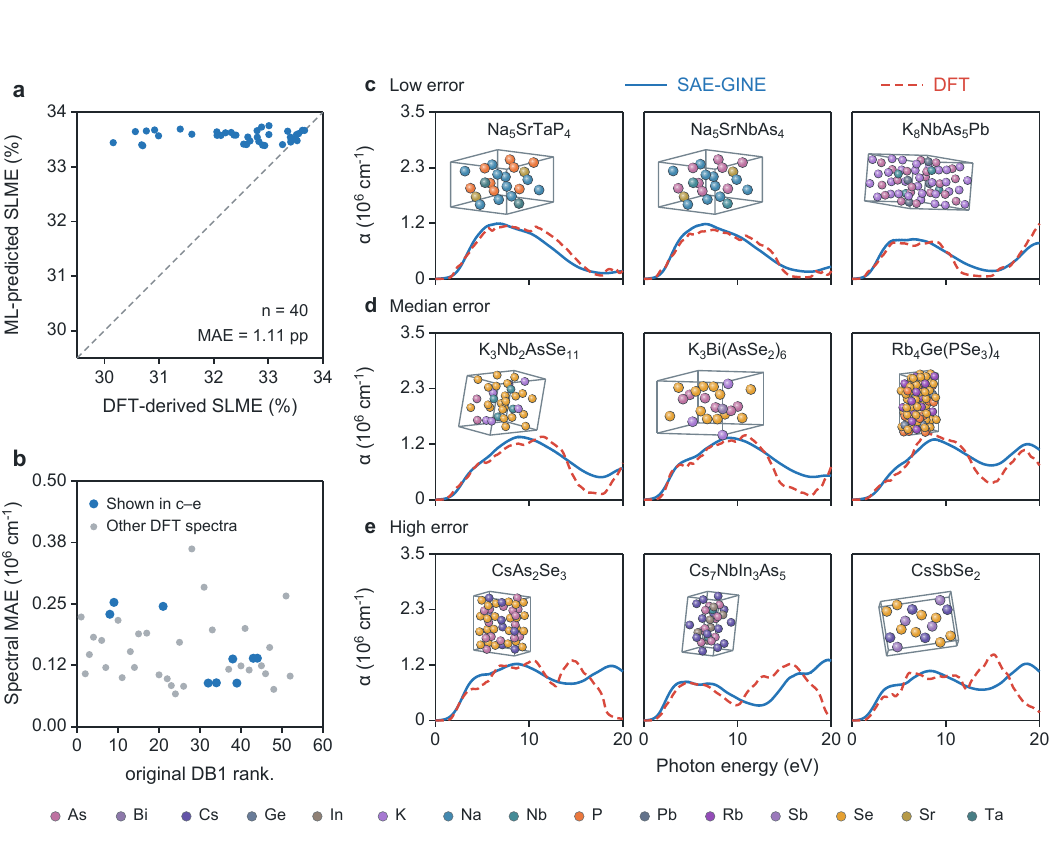}
    \caption[External DFT validation of SLME and absorption spectra]{%
    External DFT validation of SLME and representative absorption
    spectra.
    (a) ML-predicted versus DFT-derived SLME for the 40 evaluated
    candidates; the dashed line denotes ideal agreement.
    (b) Per-material raw absorption-spectrum MAE over 0-20~eV as a
    function of the original DB1 shortlist rank. Blue symbols identify
    the nine spectra shown in panels (c-e), whereas grey symbols denote
    the other DFT spectra included in the validation set.
    Panels (c), (d), and (e) show three examples nearest to the 10th,
    50th, and 90th percentiles, respectively, of the spectral-MAE
    distribution, with examples within each group ordered by increasing
    MAE. Selection is based solely on spectral MAE and is independent of
    SLME values or SLME errors.
    Solid blue and dashed red curves denote SAE-GINE predictions and
    DFT-derived absorption spectra, respectively. Crystal-structure insets
    show the corresponding candidate structures, with atomic species
    identified by the shared color key.
    The SLME comparison uses the common Materials Project band-gap cutoff
    with $L=500$~nm, $T=300$~K, and $f_r=1$, whereas the raw spectral
    comparisons are performed over 0-20~eV without applying the
    band-gap cutoff.
    }
    \label{fig:dft_validation}
\end{figure*}

\section{First-Principles Validation of Screened Candidates}
\label{sec:dft_validation}

To assess whether machine-learning screening identifies candidates with
favorable first-principles optical responses, we perform an external DFT
comparison for the Zintl-like shortlist.
Among the 50 highest-ranked candidates retained from screening, 40 are
included in the comparison between DFT and ML (see Table S10). Three candidates,
Rb$_2$ThPSe$_3$, K$_2$Sm$_2$As$_2$Se$_9$, and Rb$_3$CePSe$_4$,
were not evaluated because pseudopotentials for Th, Sm, and Ce were not
available in the pseudopotential set used in this work. Six additional
candidates, Rb$_2$HgPSe$_5$, KAgPSe$_3$, Ba$_2$GeSb$_2$S$_5$,
Ba$_4$Sb$_4$Se$_{11}$, Li$_3$As$_7$, and Cs$_4$Sb$_{14}$S$_{23}$,
were not included in the optical calculations because their large electron
counts required a very large number of Kohn-Sham bands, making the
calculations too demanding for the computational resources available in this
study. One additional candidate, Sr$_4$TiAs$_4$ (mp-wyf), is not included
in the final 40-candidate quantitative comparison. Further computational
details are provided in the Theoretical Section.

Figure~\ref{fig:dft_validation}a compares the ML-predicted and DFT-derived SLME values for the 40 evaluated candidates. Under the common screening protocol, all DFT-derived SLME values remain above 30\%, spanning
30.16-33.66\%, with an ML-DFT mean absolute error of
1.11 percentage points. Both ML- and DFT-derived SLME values are evaluated
using the same Materials Project band-gap cutoff, absorber thickness of
500~nm, temperature of 300~K, and $f_r=1$. Retention of the screening
threshold across the evaluated set supports the use of SAE-GINE for
candidate prioritization. The ML predictions nevertheless occupy a
narrower range than the DFT-derived values, indicating that agreement at
the screening threshold does not imply equally accurate resolution among
closely ranked candidates.

The absorption-spectrum comparison provides a complementary assessment
of the predicted optical response across the full energy window.
Figure~\ref{fig:dft_validation}b reports the per-material spectral MAE
over 0-20~eV for all 40 candidates, while
Figures~\ref{fig:dft_validation}c-e samples the low-, median-, and
high-error regions of this distribution. The corresponding spectral-MAE
ranges are approximately
$(0.089$-$0.089)\times10^{6}$~cm$^{-1}$,
$(0.138$-$0.140)\times10^{6}$~cm$^{-1}$, and
$(0.229$-$0.253)\times10^{6}$~cm$^{-1}$.

The low- and median-error examples reproduce the principal absorption
bands and broad energy-dependent profiles, with remaining differences
primarily in local spectral features and amplitudes. The high-error
examples show more pronounced discrepancies in both spectral shape and
amplitude. In particular, CsAs$_2$Se$_3$, Cs$_7$NbIn$_3$As$_5$, and
CsSbSe$_2$ have Pearson correlation coefficients of approximately
0.561, 0.381, and 0.439, respectively. These cases illustrate that
full-spectrum error can reflect both amplitude and shape differences,
and that correlation and absolute error provide complementary measures
of agreement. Material-resolved SLME and spectral metrics for all 40
external DFT calculations are reported in Table S10 of the Supporting
Information, and the complete raw 0-20~eV absorption-spectrum
comparisons are shown in Figure S5.

Together, the SLME and spectrum-level comparisons connect the screening
predictions to external first-principles optical calculations. The
evaluated candidates retain the target SLME threshold under the adopted
protocol, whereas the raw spectra expose differences that are less
apparent from the integrated efficiency metric alone. These results support
SAE-GINE as a tool for candidate prioritization while retaining
first-principles calculations as the quantitative validation step.
The conclusions apply to the evaluated subset, and the reported SLME
values should be interpreted as screening metrics rather than predictions
of realized device efficiency.

\section{Discussion}
\label{sec:discussion}

The results demonstrate the value of explicit chemical representation
for full-spectrum optical learning. Within the evaluated GINE framework,
attribute enrichment reduces the dielectric-spectrum MAE by 12.3\% for
\(\varepsilon_1(E)\) and 10.2\% for \(\varepsilon_2(E)\) relative to the
basic representation. Encoding physicochemical information at both
node and edge levels allows neighboring atomic pairs with similar
geometric separations to remain distinguishable during message passing.
Because the graph configurations differ jointly in descriptor content
and preprocessing, these gains establish the benefit of the complete
representation rather than any individual descriptor. Self-supervised
initialization provides additional improvements under the matched
full-data and intermediate-label conditions. Graph enrichment and
pretraining thus address complementary aspects of learning: the
information supplied to the encoder and its organization before
optical supervision.

%The methodological contribution lies in combining node-edge chemical enrichment with self-supervised structural pretraining and examining their roles through internal comparisons. The evidence supports this combination under the evaluated dataset and protocols; differences in data, architectures, and evaluation definitions limit direct numerical comparisons with other optical-learning frameworks. The predicted dielectric spectra also preserve information relevant to derived optical responses. However, the finite-window Kramers-Kr\"onig diagnostic distinguishes strong spectral-shape correspondence from quantitative consistency between independently predicted dielectric components. Joint prediction and causality-aware constraints therefore offer directions for improving this consistency.

For photovoltaic screening, the dedicated direct-\(\alpha(E)\) branch
predicts the optical response used in SLME evaluation. Across the
853 held-out cases with finite SLME values, it achieves
\(R^2=0.983\) and an MAE of 0.59 percentage points. 
%Top-50 and Top-100 retention rates of 88\% and 89\%, respectively, support its use for preserving a promising candidate pool, while the lower Top-10 retention of 50\% indicates greater sensitivity when selecting only the highest-ranked few. These results establish the utility of the direct-absorption branch for prioritization; its relative advantage over dielectric-derived absorption would require a matched head-to-head evaluation.
External screening extends this workflow to chemically targeted
Zintl-like and \(AB_2C_2\) compounds. SLME filtering,
reference-structure matching, source-pool shortlist selection, and
cross-pool deduplication reduce 1,172 eligible structures to 65 unique
candidates. Structural novelty is defined relative to the adopted
reference database and matching protocol. Likewise, the common
Materials Project band-gap cutoff and the assumption \(f_r=1\) define
a consistent radiative-limit screening protocol. The resulting SLME
values support computational prioritization but do not incorporate
material-specific nonradiative recombination or other device losses.

The DFT comparison provides an independent assessment of this
prioritization strategy for 40 candidates from the Zintl-like shortlist.
All 40 retain DFT-derived SLME values above 30\% under the common
screening protocol, with an ML-DFT MAE of 1.11 percentage points.
The narrower spread of the ML-predicted values, however, shows that
retaining the screening threshold is distinct from accurately resolving
differences among closely ranked candidates.

The raw absorption comparisons over 0-20~eV reinforce this distinction.
Low- and median-error examples reproduce the principal absorption bands
and broad spectral profiles, whereas high-error examples exhibit
differences in both amplitude and spectral shape, particularly at
higher energies. The integrated SLME metric and the spectrum-level
comparison therefore probe complementary aspects of model performance:
the former evaluates screening-level retention, while the latter exposes
optical-response discrepancies that can be obscured by a single efficiency
score.

Overall, these results connect chemically informed crystal
representation, optical-spectrum prediction, and photovoltaic
candidate prioritization. Attribute enrichment improves the
evaluated structure-to-spectrum mapping, while self-supervised
initialization provides complementary gains under selected
training conditions. The direct-absorption branch translates
these capabilities into a physically motivated screening
workflow, with targeted first-principles calculations assessing
the shortlisted candidates and resolving material-specific
optical differences.

\section{Experimental Section}
\label{sec:experimental}

\methodhead{Dataset of Crystal Structure – Optical Spectrum}

The database contains 10,191 crystal records with optical data \cite{Grunert2024}, which are calculated within the independent-particle approximation (IPA). 
%After excluding two entries without usable scalar or spectral information, 10,189 valid records remain. 
The keys \texttt{epsR\_0} and \texttt{epsI\_0} are used as the real and imaginary dielectric responses, \(\varepsilon_1(E)\) and \(\varepsilon_2(E)\), respectively.
%The suffix \texttt{0} is retained as a dataset-defined component label and is not assigned to a specific Cartesian direction. 
Both dielectric components are required for each retained crystal so that the dispersive and absorptive responses originate from the same periodic structure. The final paired dataset contains 9,915 crystal-spectrum samples. Each spectrum contains 2001 values on a common photon-energy grid spanning 0-20~eV.

Before model development, the paired dielectric views are verified to have identical crystal indices, spectral dimensions, and photon-energy grids. A fixed record-level partition generated using seed 2025 contains 7,931 training, 992 validation, and 992 test structures and is used consistently for both dielectric components and all principal model comparisons (see Table S1). No composition- or structural-prototype grouping is imposed in the primary split. An overlap audit confirms that no exact base record, material identifier, or source filename is shared across the training, validation, and test partitions. Under the \texttt{StructureMatcher} \cite{Ong2013} tolerances specified in the Supporting Information, structure-equivalent counterparts in the training partition are identified for 13 validation and 12 test structures, corresponding to 1.31\% and 1.21\% of the respective sets. The benchmark is therefore interpreted as a fixed record-level random-split evaluation rather than a strict chemical or structural out-of-distribution test. The detailed structure-overlap analysis is reported in the
Supporting Information.

To evaluate optical-label efficiency, the fixed training partition is additionally subsampled to \(5\%\), \(10\%\), \(25\%\), and \(50\%\), corresponding to 397, 793, 1,983, and 3,966 labeled structures, respectively, while the validation and test partitions remain unchanged. Within each fraction, AE-GINE and SAE-GINE use identical labeled structures. The primary subsets are generated using seed 777 and are sampled independently across fractions.

\methodhead{Crystal-Graph Representation}

Each periodic crystal is represented as a directed graph in which atomic sites define the nodes and periodic neighboring pairs define the edges. Neighboring sites are identified within a radial cutoff of \(5.0~\text{\AA}\), and up to the 16 nearest neighbors are retained for each central site before reverse-edge completion.
Zero-distance self-connections in the reference unit cell are removed, whereas
connections involving periodic images are preserved. 
%Missing reverse connections are subsequently added to permit bidirectional information exchange. 
The same crystal structures and neighbor topology are retained across all graph representations so that differences between the basic and attribute-enriched descriptions arise from the
encoded node and edge information rather than from changes in graph connectivity.

The basic representation contains 11 node features and 16 edge features. Each atomic
site is described by atomic number, periodic-table row and group, atomic mass,
electronegativity, atomic radius, and first ionization energy, together with four
binary indicators recording the availability of atomic mass, electronegativity,
atomic radius, and ionization energy. Missing continuous values are replaced by zero
and distinguished through the corresponding availability indicators. Structural
information on each edge is represented by expanding the periodic interatomic distance
\(d_{ij}\) over 16 Gaussian radial-basis functions \cite{Schtt2018,Unke2019},
\begin{equation}
\phi_m(d_{ij})=
\exp\left[-\gamma(d_{ij}-\mu_m)^2\right],
\label{eq:rbf}
\end{equation}
where the centers \(\mu_m\) are uniformly distributed between 0 and
\(5.0~\text{\AA}\), and \(\gamma=(\Delta\mu)^{-2}\).

The primary attribute-enriched representation contains 19 node features and 24 edge
features and is used throughout the dielectric-spectrum benchmarks (see Tables S2 and S3), self-supervised
pretraining, limited-label experiments, ensemble prediction, direct-\(\alpha(E)\)
modeling, and external screening. Relative to the basic representation, the node
descriptor set is redesigned to incorporate additional chemically informative
quantities, including calculated atomic-radius and ionic-radius descriptors,
valence-shell occupations, periodic-table block, and metallic-character descriptors,
while maintaining a fixed 19-dimensional enriched-node representation. Continuous
elemental descriptors are scaled using fixed descriptor-specific factors that are
independent of the crystal-spectrum dataset. The complete definitions of the
19 node-features and their preprocessing and missing-value conventions are
provided in the Supporting Information.

The enriched edge representation retains the 16 radial channels and supplements them
with eight pairwise physicochemical attributes,
\begin{equation}
\begin{aligned}
\mathbf{e}_{ij}^{\mathrm{enriched}}=\big[&
\boldsymbol{\phi}(d_{ij}),
|\chi_i-\chi_j|,
|r_i-r_j|,\\
& r_i+r_j,
|I_i-I_j|,
|Z_i-Z_j|,\\
& \delta_{\mathrm{element}},
\delta_{\mathrm{group}},
\delta_{\mathrm{row}}\big],
\end{aligned}
\label{eq:enriched_edge}
\end{equation}
where \(\boldsymbol{\phi}(d_{ij})
=[\phi_1(d_{ij}),\ldots,\phi_{16}(d_{ij})]\) denotes the 16-dimensional
radial-basis expansion, while \(\chi\), \(r\), \(I\), and \(Z\) denote
electronegativity, atomic radius, first ionization energy, and atomic number,
respectively. The binary terms indicate whether the two atoms in a connected pair
are the same element or belong to the same periodic-table group or row. The enriched
representation therefore preserves periodic geometry while explicitly encoding
physicochemical contrasts between connected atomic sites. The precise construction
and preprocessing conventions for the eight pairwise attributes are provided together
with the descriptor specification in the Supporting Information.

Elemental descriptors are obtained from the \texttt{Element} class in
\texttt{pymatgen.core.periodic\_table} \cite{Ong2013}, and missing continuous values are replaced
by zero. Because the basic and attribute-enriched representations differ jointly in
descriptor content, dimensionality, numerical scaling, missing-value treatment, and
pairwise chemical information, their comparison is interpreted as an evaluation of
the complete graph representations rather than as an isolated single-feature
ablation.

\methodhead{Self-Supervised Pretraining and Spectrum Prediction}

Self-supervised pretraining uses only the 7,931 attribute-enriched crystal graphs in the fixed training partition. Validation and test structures are excluded, and no dielectric-spectrum, band-gap, derived-optical, or photovoltaic target enters the pretraining objective. The pretext task perturbs the node and edge attributes while preserving the periodic graph topology. Node-feature vectors are independently masked with probability 0.15, followed by zero-mean Gaussian perturbations with a standard deviation of 0.02 applied to the node and edge feature channels,
\begin{equation}
\widetilde{\mathbf{x}}_i=
(1-m_i)\mathbf{x}_i+\boldsymbol{\eta}_i,
\qquad
\widetilde{\mathbf{e}}_{ij}=
\mathbf{e}_{ij}+\boldsymbol{\xi}_{ij},
\label{eq:ssl_corruption}
\end{equation}
where \(m_i\sim\mathrm{Bernoulli}(0.15)\), and
\(\boldsymbol{\eta}_i\) and \(\boldsymbol{\xi}_{ij}\) denote Gaussian perturbations with a standard deviation of 0.02. Because the perturbation is applied after masking, a masked node retains only the noise contribution. Edge indices, periodic-image information, and graph connectivity remain unchanged.

The reconstruction target is calculated from the original uncorrupted graph. For each crystal, the channel-wise means and standard deviations of the node and edge attributes are concatenated as
\begin{equation}
\mathbf{t}^{\mathrm{SSL}}_G=
[
\boldsymbol{\mu}_{x},
\boldsymbol{\sigma}_{x},
\boldsymbol{\mu}_{e},
\boldsymbol{\sigma}_{e}
].
\label{eq:ssl_target}
\end{equation}
For the representation with 19 node-feature channels and 24 edge-feature channels, this produces an 86-dimensional graph-level target. Variances are bounded below by \(10^{-8}\) before evaluation of the standard deviations. The pretext task therefore reconstructs graph-level chemical and local-environment statistics from corrupted attributes rather than reconstructing individual masked atoms or using optical labels.

The corrupted graphs are processed by a four-layer attribute-enriched GINE encoder with a hidden dimension of 192 \cite{xu2018powerful,hu2019strategies}. Residual connections, layer normalization, GELU activation, and a dropout probability of 0.10 are used in the encoder \cite{He2016,ba2016layer, Hendrycks2023}, while the 24-dimensional edge attributes enter the edge-aware message-passing operations. Global mean pooling produces a crystal-level representation, which is mapped to the 86-dimensional self-supervised target through a reconstruction head containing two 256-dimensional hidden layers. Pretraining minimizes the mean Smooth-L1 loss \cite{Huber1964} with \(\beta=1.0\) for 80 epochs using AdamW \cite{loshchilov2017decoupled}, a batch size of 32, an initial learning rate of \(1\times10^{-3}\), a weight decay of \(1\times10^{-5}\), cosine learning-rate scheduling \cite{loshchilov2016sgdr}, and gradient clipping at 5.0. No separate validation subset is introduced for the self-supervised objective; the checkpoint with the lowest epoch-averaged reconstruction loss is retained.

For downstream transfer, we remove the graph-statistics reconstruction head and retain the pretrained GINE backbone. The transferred backbone includes the node encoder, four GINE layers with their edge projections, residual blocks, and normalization parameters. Separate prediction heads are newly initialized for \(\varepsilon_1(E)\) and \(\varepsilon_2(E)\), and all transferred parameters remain trainable during supervised fine-tuning.

Each supervised model maps a periodic crystal graph to a 2,001-point dielectric spectrum on the common 0-20~eV photon-energy grid. The 19-dimensional node representation is projected to a 192-dimensional hidden space and propagated through four edge-aware GINE layers using the 24-dimensional edge attributes. Global mean pooling yields the crystal representation, which is mapped to the target spectrum through two 256-dimensional hidden layers and a final 2,001-dimensional output layer. The GCN \cite{kipf2016semi} and GraphSAGE \cite{NIPS2017_5dd9db5e} baselines use the same hidden dimension, number of message-passing layers, pooling operation, and output dimensionality, but do not use the stored edge attributes during message passing. Their comparison with GINE therefore evaluates the complete implemented graph-learning configurations rather than the aggregation operators in isolation.

To account for the different numerical scales across photon energy, the dielectric spectra are standardized channel-wise using statistics calculated only from the corresponding supervised training set,
\begin{equation}
z_{G,c}=
\frac{y_{G,c}-\mu_c}
{\sigma_c+\epsilon_{\mathrm{num}}},
\qquad
\widehat{z}_{G,c}=
\frac{\widehat{y}_{G,c}-\mu_c}
{\sigma_c+\epsilon_{\mathrm{num}}},
\label{eq:spectral_normalization}
\end{equation}
where \(\mu_c\) and \(\sigma_c\) denote the training-set mean and population standard deviation at energy channel \(c\), and \(\epsilon_{\mathrm{num}}=10^{-8}\). For the limited-label experiments, these statistics are recalculated using only the retained labeled structures at the corresponding fraction; validation and test spectra do not contribute to the normalization.

The supervised objective combines a peak-weighted Smooth-L1 reconstruction term with a first-difference constraint and, for \(\varepsilon_2(E)\), a soft penalty on negative predictions. The spectral weight is defined as
\begin{equation}
w_{G,c}=
1+
\lambda_{\mathrm{peak}}
\frac{|y_{G,c}|}
{\max_{c'}|y_{G,c'}|+\epsilon_{\mathrm{num}}},
\qquad
\lambda_{\mathrm{peak}}=2.0,
\label{eq:peak_weight}
\end{equation}
so that stronger dielectric responses receive greater weight while all energy channels remain represented. The total supervised objective is
\begin{equation}
\mathcal{L}_{\mathrm{sup}}=
\mathcal{L}_{\mathrm{spec}}
+
\lambda_{\Delta}\mathcal{L}_{\Delta}
+
\lambda_{\mathrm{neg}}\mathcal{L}_{\mathrm{neg}},
\label{eq:supervised_loss}
\end{equation}
with \(\lambda_{\Delta}=0.03\). The local spectral term compares adjacent predicted and reference values and is treated as a first-difference constraint rather than as a numerical derivative. The negative-response coefficient is set to zero for \(\varepsilon_1(E)\) and to 0.02 for \(\varepsilon_2(E)\).

Each supervised configuration is evaluated using seeds 42, 123, and 2025, a batch size of 16, and a maximum of 100 epochs. Optimization uses AdamW, gradient clipping at 5.0, and automatic mixed precision. A reduce-on-plateau scheduler monitors validation MAE with a reduction factor of 0.5, a patience of four epochs, and a minimum learning rate of \(10^{-6}\). Training terminates after 15 epochs without validation improvement, and the checkpoint with the lowest validation MAE is retained. The fixed test partition is evaluated only after model selection.

The full-data configurations in Table~\ref{tab:main_benchmark} represent the final evaluated settings rather than a strictly controlled initialization ablation. AE-GINE uses an initial learning rate of \(5\times10^{-4}\), a weight decay of \(1\times10^{-4}\), and a dropout probability of 0.10, whereas SAE-GINE uses \(3\times10^{-4}\), \(5\times10^{-5}\), and 0.05, respectively. The matched full-data comparison in Table~\ref{tab:matched_full_data} instead keeps the graph representation, data partitions, architecture, optimization settings, and training seeds identical so that encoder initialization is the only controlled difference. In the limited-label experiments, both initialization strategies use a learning rate of \(3\times10^{-4}\), a weight decay of \(5\times10^{-5}\), and a dropout probability of 0.05 at every evaluated label fraction.

\methodhead{Metric Parameters to Evaluate the Optical Prediction}

Model performance is evaluated after transforming the predicted dielectric spectra back
to their original numerical scale. For each training run, the fixed test partition is
accessed only after the checkpoint with the lowest validation mean absolute error (MAE) is selected. For test structure \(i\), the spectrum-level MAE is defined as
\begin{equation}
\mathrm{MAE}_i=
\frac{\sum_{c=1}^{C}m_{i,c}|\widehat{y}_{i,c}-y_{i,c}|}
{\sum_{c=1}^{C}m_{i,c}},
\label{eq:mae}
\end{equation}
where \(C=2001\) and \(m_{i,c}\) denotes the validity mask. Then, the average MAE for the entire test set is given by
\begin{equation}
\mathrm{MAE}=
\frac{1}{N_{\mathrm{test}}}
\sum_{i=1}^{N_{\mathrm{test}}}\mathrm{MAE}_i.
\label{eq:mae1}
\end{equation}
All paired dielectric spectra retained in this work contain complete targets, so the reported MAE is equivalent to the mean absolute pointwise error across the held-out structures and the full photon-energy grid.

%We also evaluate spectral reconstruction using a weighted coefficient, which is defined as
%\begin{equation}
%R_{\mathrm{w}}^2=
%1-
%\frac{\sum_{c\in\mathcal{V}}\mathrm{SSE}_c}
%{\sum_{c\in\mathcal{V}}\mathrm{SST}_c},
%~~
%\mathcal{V}=
%\left\{
%c:\mathrm{SST}_c>10^{-12}
%\right\},
%\label{eq:weighted_r2}
%\end{equation}
%where the sum of squared errors (SSE) and the sum of squares total (SST) are given for each energy point $c$ as:
%\begin{align}
%\overline{y}_c&=
%\frac{\sum_i m_{i,c}y_{i,c}}
%{\sum_i m_{i,c}},\\
%\mathrm{SSE}_c&=
%\sum_i m_{i,c}(y_{i,c}-\widehat{y}_{i,c})^2,\\
%\mathrm{SST}_c&=
%\sum_i m_{i,c}(y_{i,c}-\overline{y}_c)^2.
%\end{align}
%This weighting gives greater contribution to photon-energy channels that vary across materials while limiting the influence of nearly invariant channels. Each model configuration is evaluated over three independent training seeds. MAE values are reported as the mean \(\pm\) sample standard deviation calculated with the \(S-1\) denominator, whereas \(R_{\mathrm{w}}^2\) is evaluated for each run and summarized across the three runs. Error bars in the limited-label analysis likewise denote one sample standard deviation.

Prediction-level ensembles are constructed by averaging the denormalized predictions
of the three independently trained models pointwise,
\begin{equation}
\widehat{y}^{\mathrm{ens}}_{i,c}=
\frac{1}{3}
\sum_{s=1}^{3}
\widehat{y}^{(s)}_{i,c}.
\label{eq:ensemble}
\end{equation}
The ensemble therefore combines model predictions rather than network parameters.
Before averaging, the reference spectra, photon-energy grids, and stored sample indices
are verified for consistency across the three runs. The component-specific configurations used for held-out downstream analysis
are selected using the three-run mean validation MAE without reference to
held-out test performance. This procedure selects AE-GINE for
\(\varepsilon_1(E)\) and SAE-GINE for \(\varepsilon_2(E)\).
Each prediction-level ensemble contains seeds 42, 123, and 2025, with each
run represented by the checkpoint attaining its lowest validation MAE.

%For the cross-response comparison in Figure~\ref{fig:optical_r2}, Equation~\eqref{eq:weighted_r2} is applied to the final prediction-level ensemble outputs and to the optical responses derived from the same dielectric predictions. The directly predicted \(\varepsilon_1(E)\) and \(\varepsilon_2(E)\) responses and the derived \(n(E)\), \(k(E)\), and \(R(E)\) responses use all 2,001 photon-energy points. We note that \(\alpha(E)\) at \(E=0\) has \(\mathrm{SST}_c=0\) and thus it is excluded automatically by the predefined validity set \(\mathcal{V}\), leaving 2,000 points.

Since \(\varepsilon_1(E)\) and \(\varepsilon_2(E)\) are predicted independently, their mutual consistency is assessed post hoc using the Kramers-Kr\"onig relations~\cite{Hung2024},
\begin{equation}
    \varepsilon_1(E)-1=
\frac{2}{\pi}\,
\mathcal{P}\!\int_0^\infty
\frac{E'\varepsilon_2(E')}
{E'^2-E^2}\,dE',
\label{eq:kk1}
\end{equation}
and
\begin{equation}
\varepsilon_2(E)=
-\frac{2E}{\pi}\,
\mathcal{P}\!\int_0^\infty
\frac{\varepsilon_1(E')-1}
{E'^2-E^2}\,dE'.
\label{eq:kk2}
\end{equation}
The numerical diagnostic uses trapezoidal quadrature on the available 0-20~eV grid and omits the singular \(E'=E\) term under the principal-value prescription. The same finite-window operator is applied to reference and predicted spectra so that model-induced inconsistency can be compared with the truncation residual already present in the reference data. 
%Primary residuals are evaluated over 1-18~eV, with 0.5-18, 1-17, and 2-17~eV used as window-sensitivity checks. 
Because the formal Kramers-Kr\"onig relations \cite{Toll1956,kuzmenko2005kramers} require integration over an infinite frequency range, these finite-window residuals are interpreted as a diagnostic rather than as a direct test of exact causality.

Representative structures in Figure~\ref{fig:representative_spectra} are selected
from the distribution of per-structure ensemble MAEs rather than by visual inspection.
If \(Q_p\) denotes the \(p\)th percentile of this distribution,
\begin{equation}
i_p=
\underset{i}{\arg\min}
\left|
\mathrm{MAE}^{\mathrm{ens}}_i-Q_p
\right|,
\qquad
p\in\{10,50,90\}.
\label{eq:percentile_selection}
\end{equation}
The resulting examples therefore correspond to actual held-out structures nearest to the low-error, median-error, and high-error percentiles of the ensemble-error distribution.

%The square-root arguments are bounded below by zero for numerical stability,
%\(\alpha(E)\) is reported in \(\mathrm{cm}^{-1}\), and \(R(E)\) is restricted to the interval \([0,1]\). These dielectric-derived quantities are used to evaluate how errors in \(\varepsilon_1(E)\) and \(\varepsilon_2(E)\) propagate to related optical responses; they are not used to construct the final photovoltaic shortlist.

\methodhead{Photovoltaic Materials Screening}

For photovoltaic screening, we use a separate SAE-GINE model that predicts the
absorption-coefficient spectrum \(\alpha(E)\) directly from crystal structure. The reference \(\alpha(E)\) targets are constructed from the paired \(\varepsilon_1(E)\) and \(\varepsilon_2(E)\) spectra using the dielectric-to-absorption relations defined above and are expressed in unit of \(\mathrm{cm}^{-1}\) as follows \cite{hung2022quantum}:
\begin{equation}
\alpha(E)=\frac{2E}{\hbar c}\sqrt{\frac{\sqrt{\varepsilon_1(E)+\varepsilon_2(E)}-\varepsilon_1(E)}{2}}.
\label{eq:optical_transformations}
\end{equation}

The dielectric spectra are used only to construct the supervised absorption target; the downstream model receives the crystal graph alone and therefore predicts \(\alpha(E)\) without first predicting \(\varepsilon_1(E)\) or \(\varepsilon_2(E)\). The supervised target is
\begin{equation}
y_{\alpha}(E)=
\log_{10}\left[1+\alpha(E)\right].
\label{eq:direct_alpha_target}
\end{equation}
Here, the direct-\(\alpha(E)\) task uses the same 9,915 structures and fixed record-level partition as the dielectric models, comprising 7,931 training, 992 validation, and 992 test structures. The transformed target contains 2,001 energy values on the common 0-20~eV grid and is standardized value-wise using statistics calculated only from the training partition.

The direct-\(\alpha(E)\) SAE-GINE uses the primary attribute-enriched graph
representation with 19-dimensional node features and 24-dimensional edge
features and the same \texttt{OpticalResponseGINE} architecture used for the
dielectric-spectrum models. The node features are projected to a hidden dimension
of 192 and propagated through four edge-aware GINE layers, followed by global mean
pooling and a spectral prediction head containing two 256-dimensional hidden layers.
The backbone is initialized from the self-supervised checkpoint described above, and
all transferred parameters remain trainable during supervised fine-tuning.

Three independent models are trained using seeds 42, 123, and 2025. Training uses
AdamW with an initial learning rate of \(3\times10^{-4}\), a weight decay of
\(5\times10^{-5}\), a dropout probability of 0.05, a batch size of 16, and a maximum of 100 epochs. The supervised objective follows the same peak-weighted Smooth-L1 and first-difference formulation used for the dielectric models, with
\(\lambda_{\mathrm{peak}}=2.0\) and \(\lambda_{\Delta}=0.03\). For this task,
peak weighting and the negative-response constraint are applied in the denormalized transformed-target space \(y_{\alpha}(E)\). Because
\(y_{\alpha}(E)=\log_{10}[1+\alpha(E)]\geq0\) for non-negative absorption
coefficients, predictions below zero are softly penalized using
\(\lambda_{\mathrm{neg}}=0.02\). Gradient clipping at 5.0 and automatic
mixed-precision training are used. The learning rate is reduced by a factor of 0.5
after four epochs without improvement in validation MAE, with a minimum learning
rate of \(10^{-6}\), and training terminates after 15 epochs without validation
improvement. For each seed, the checkpoint attaining the lowest validation MAE is
retained.

After inverse standardization, the three retained models are combined by averaging
their predictions in the transformed target space \(y_{\alpha}(E)\),
\begin{equation}
\widehat{y}_{\alpha}^{\mathrm{ens}}(E)=
\frac{1}{3}
\sum_{s=1}^{3}
\widehat{y}_{\alpha}^{(s)}(E).
\label{eq:direct_alpha_ensemble}
\end{equation}
The physical absorption coefficient is then recovered as
\begin{equation}
\widehat{\alpha}(E)=
10^{\widehat{y}_{\alpha}^{\mathrm{ens}}(E)}-1.
\label{eq:direct_alpha_inverse}
\end{equation}

The spectroscopic limited maximum efficiency (SLME) is calculated from the absorption spectrum using the detailed-balance formulation \cite{Yu2012}. For absorber thickness \(L\), the energy-dependent absorptivity is
\begin{equation}
A(E)=
1-\exp[-2\alpha(E)L].
\label{eq:slme_absorptivity}
\end{equation}
The short-circuit current density and radiative saturation current density are
\begin{align}
J_{\mathrm{sc}}&=
q\int
A(E)\,
I_{\mathrm{solar}}(E)\,dE,
\\
J_0&=
\frac{q\pi}{f_{\mathrm r}}
\int
A(E)\,
I_{\mathrm{bb}}(E,T)\,dE,
\label{eq:slme_currents}
\end{align}
where \(I_{\mathrm{solar}}(E)\) denotes the incident AM1.5G solar photon flux
density and \(I_{\mathrm{bb}}(E,T)\) denotes the blackbody photon radiance at
temperature \(T\). The current density at voltage \(V\) follows the ideal-diode
relation \cite{Shockley1961}
\begin{equation}
J(V)=
J_{\mathrm{sc}}
-
J_0
\left[
\exp\left(
\frac{qV}{k_{\mathrm B}T}
\right)-1
\right].
\label{eq:slme_diode}
\end{equation}
The maximum output power density is
\begin{equation}
P_{\max}=
\max_V\left[J(V)V\right],
\end{equation}
while the incident solar power density is
\begin{equation}
P_{\mathrm{solar}}=
\int
E\,I_{\mathrm{solar}}(E)\,dE.
\end{equation}
The resulting SLME is given by
\begin{equation}
\eta_{\mathrm{SLME}}=
\frac{P_{\max}}{P_{\mathrm{solar}}}.
\label{eq:slme}
\end{equation}

All SLME calculations use the ASTM G173 AM1.5G spectrum \cite{G03_Committee2020-wo},
\(L=500~\mathrm{nm}\), \(T=300~\mathrm{K}\), and \(f_{\mathrm r}=1\).
For held-out test evaluation, \(\alpha(E)\) is set to zero below the independently
available direct band gap for both the reference and predicted spectra before SLME
evaluation. The same gap cutoff is therefore applied to both sides of the held-out
comparison. For the external candidate pools, independent direct-gap labels are
unavailable, and the Materials Project band gap is instead used as the common
numerical cutoff. The same external-screening cutoff convention is retained in the
subsequent DFT comparison so that ML- and DFT-derived SLME values are evaluated
under a common screening protocol.

\methodhead{Candidate Screening and DFT Shortlist Construction}

External crystal structures and electronic-structure metadata are
obtained from the Materials Project \cite{Jain2013}. Two complementary
search spaces are considered: a compositionally defined Zintl-like
pool and a ternary \(AB_2C_2\) pool with \(C=\mathrm{P}\) or
\(\mathrm{N}\).

The Zintl-like search requires at least one donor element from
\(\{\mathrm{Li},\mathrm{Na},\mathrm{K},\mathrm{Rb},\mathrm{Cs},
\mathrm{Mg},\mathrm{Ca},\mathrm{Sr},\mathrm{Ba},\mathrm{Eu},
\mathrm{Yb}\}\)
and at least one framework element from
\(\{\mathrm{B},\mathrm{Al},\mathrm{Ga},\mathrm{In},\mathrm{Tl},
\mathrm{Si},\mathrm{Ge},\mathrm{Sn},\mathrm{Pb},\mathrm{P},
\mathrm{As},\mathrm{Sb},\mathrm{Bi}\}\).
For this pool, compositions containing H, C, N, O, halogens,
or noble gases are excluded. Compositions must also admit an
oxidation-state assignment with a positively charged donor and
a negatively charged framework element. These criteria define
a chemically motivated Zintl-like search space
\cite{Kauzlarich2023,Fang2022,Chaliha2025}, without establishing
every retained structure as a Zintl phase. The \(AB_2C_2\) pool
is constructed separately from reduced ternary compositions with
a \(1{:}2{:}2\) stoichiometric ratio and
\(C=\mathrm{P}\) or \(\mathrm{N}\).

For both pools, eligibility requires a Materials Project band gap
of 0.30-4.50~eV and an energy above the convex hull of at most
\(0.10~\mathrm{eV\,atom^{-1}}\) \cite{Sun2016,Bartel2020}.
Duplicate material identifiers are removed within each pool.
The Materials Project stability flag is retained as metadata
rather than used as an additional exclusion criterion.
The initial searches contain 1,361 Zintl-like and 285 \(AB_2C_2\)
structures, of which 1,100 and 72, respectively, satisfy the
eligibility criteria.

Eligible structures are converted into the primary
attribute-enriched graph representation with 19-dimensional node
features and 24-dimensional edge features. The three-seed
direct-\(\alpha(E)\) SAE-GINE ensemble, trained with seeds
42, 123, and 2025, is applied to each structure. After inverse
standardization, predictions are averaged in the
\(\log_{10}[1+\alpha(E)]\) target space and transformed back to
the physical absorption coefficient using
Equations~\eqref{eq:direct_alpha_target}-%
\eqref{eq:direct_alpha_inverse}. SLME is calculated using
Equations~\eqref{eq:slme_absorptivity}-\eqref{eq:slme}
with \(L=500\)~nm, \(T=300\)~K, and \(f_r=1\).
Because independent direct-gap labels are unavailable for these
external pools, the Materials Project band gap is used as a common
numerical cutoff, with \(\alpha(E)\) set to zero below this energy
before SLME evaluation.

The primary criterion, predicted \(\mathrm{SLME}>30\%\), retains
355 Zintl-like and 17 \(AB_2C_2\) structures. These candidates are
compared with the original crystal-spectrum database using the
adopted structure-matching protocol, and structure-equivalent
entries are removed. Reduced-formula coincidence alone is
insufficient for exclusion because distinct polymorphs can share
the same composition. After structure-overlap removal, 325
Zintl-like and 16 \(AB_2C_2\) candidates remain.

The 50 highest-ranked remaining Zintl-like candidates are selected
according to predicted SLME, together with all 16 remaining
\(AB_2C_2\) candidates. The resulting 66 source-pool memberships
include one shared structure, Ba(GaP)$_2$ (\texttt{mp-ckxsa}),
yielding 65 unique candidates with predicted SLME values of
30.16-33.75\%. This set constitutes the pre-DFT shortlist for
targeted first-principles evaluation.

\methodhead{First-Principles Calculations}

First-principles calculations are performed using Quantum ESPRESSO
(version 7.3.1) \cite{Giannozzi2009} with the
Perdew-Burke-Ernzerhof exchange-correlation functional
\cite{Perdew1996}. Crystal structures obtained from the Materials
Project are used as supplied, without additional structural relaxation
in the present workflow. Optimized norm-conserving Vanderbilt
pseudopotentials \cite{Hamann2013} are taken from the SG15 library
\cite{Schlipf2015}. Plane-wave kinetic-energy and charge-density
cutoffs of 80 and 320~Ry, respectively, are used. Calculations
are performed without spin polarization or spin-orbit coupling. The self-consistent-field (SCF) and non-self-consistent-field (NSCF)
$k$-point meshes are generated using target reciprocal-space
spacings of 0.19~\AA$^{-1}$ and 0.10~\AA$^{-1}$, respectively. Fixed
occupations are used, and the SCF convergence threshold is set
to $10^{-7}$~Ry. For the evaluated candidates, the number of
NSCF bands equals the total number of valence electrons
represented by the pseudopotentials, corresponding to twice
the number of fully occupied bands in the non-spin-polarized
calculation.

Optical responses are calculated within the independent-particle
approximation \cite{Adler1962,Wiser1963,hung2022quantum}. The dielectric spectra are evaluated at
2001 uniformly spaced energy points from 0 to 20~eV, with
Gaussian interband broadening of 0.30~eV
and no rigid energy shift. The three diagonal components of
the real and imaginary dielectric responses are averaged
separately before conversion to the scalar absorption
coefficient. The resulting absorption spectra are linearly
interpolated onto the common 2001-point ML energy grid
spanning 0-20~eV before comparison. Spectral MAE and Pearson correlation are evaluated over
0-20~eV without a band-gap cutoff, additional smoothing,
or amplitude normalization. For SLME evaluation, the same
Materials Project band-gap cutoff is applied to both ML
and DFT absorption spectra, with $L=500$~nm, $T=300$~K,
and $f_r=1$.

\medskip

\noindent\textbf{Supplementary Information} \par
\noindent Additional details on the dataset and graph attributes, auxiliary feature-weighting analysis, label-efficiency experiments,
Kramers-Kr\"onig consistency, direct-absorption/SLME validation,
external screening, and first-principles validation results.

% Acknowledgements
\medskip
\noindent\textbf{Acknowledgements} \par %delete if not applicable))
\noindent We thank Harry A. Atwater and Peter Chen for helpful discussions on Zintl-like materials. N.T.H. acknowledges the financial support from the Frontier Research Institute for Interdisciplinary Sciences from Tohoku University and the Catalyst for Change Program of the University Academic Alliance in Taiwan (UAAT), Ministry of Education, Taiwan (UAAT-CFC-11401). M.L. acknowledges the support of National Science Foundation (award no. ITE-2345084) and the support from R. Wachnik.

% Acknowledgements
\medskip
\noindent\textbf{Author Contributions} \par
\noindent Q.T.T.B and N.T.H. conceived the project. Q.T.T.B and N.T.H. developed the machine-learning modeling workflow and performed model training. N.T.G.B. carried out the first-principles calculations. M.L. and V.T.H.T. developed the concept and machine-learning modeling workflow. V.T.H.T. and N.T.H. supervised the project. All authors contributed to preparing and revising the manuscript.

\medskip
\noindent\textbf{Conflicts of Interest} \par %delete if not applicable))
\noindent The authors declare they have no conflict of interest.

\medskip
\noindent\textbf{Data Availability Statement} \par %delete if not applicable))
\noindent The data that support the findings of this study are openly available on GitHub at \url{https://github.com/nguyen-group/SAE-GINE}.

% References
\medskip

\bibliographystyle{unsrtnat}
%\IfFileExists{references.bib}{\bibliography{references}}{\par\noindent\textbf{References}\par\textit{The original references.bib file has not been supplied. Add it to compile the complete bibliography.}}

%\printbibliography
\end{document}